\documentclass[12pt,preprint]{aastex}

\shorttitle{The Cause of Supra-Arcade Downflows} 
\shortauthors{Cassak et al.}

\begin{document}

\title{On the Cause of Supra-Arcade Downflows in Solar Flares}


\author{P.~A.~Cassak\altaffilmark{1}, J.~F.~Drake\altaffilmark{2},
  J.~T.~Gosling\altaffilmark{3}, T.-D.~Phan\altaffilmark{4},
  M.~A.~Shay\altaffilmark{5}, and L.~S.~Shepherd\altaffilmark{1}}

\altaffiltext{1}{Department of Physics and Astronomy, West Virginia
  University, Morgantown, WV 26506, USA; Paul.Cassak@mail.wvu.edu,
  lshephe1@mix.wvu.edu}

\altaffiltext{2}{Department of Physics and the Institute for Physical
  Science and Technology and the Institute for Research in Electronics
  and Applied Physics, University of Maryland, College Park, MD 20742,
  USA; drake@umd.edu}

\altaffiltext{3}{Laboratory for Atmospheric and Space Physics,
  University of Colorado at Boulder, Boulder, CO 80303, USA;
  Jack.Gosling@lasp.colorado.edu}

\altaffiltext{4}{Space Science Laboratory, University of California,
  Berkeley, CA 94720, USA; phan@ssl.berkeley.edu}

\altaffiltext{5}{Department of Physics and Astronomy, University of
  Delaware, Newark, DE 20742, USA; shay@udel.edu}

\begin{abstract}
  A model of supra-arcade downflows (SADs), dark low density regions
  also known as tadpoles that propagate sunward during solar flares,
  is presented.  It is argued that the regions of low density are flow
  channels carved by sunward-directed outflow jets from reconnection.
  The solar corona is stratified, so the flare site is populated by a
  lower density plasma than that in the underlying arcade.  As the
  jets penetrate the arcade, they carve out regions of depleted plasma
  density which appear as SADs.
  The present interpretation differs from previous models in that
  reconnection is localized in space but not in time.  Reconnection is
  continuous in time to explain why SADs are not filled in from behind
  as they would if they were caused by isolated descending flux tubes
  or the wakes behind them due to temporally bursty reconnection.
  Reconnection is localized in space because outflow jets in standard
  two-dimensional reconnection models expand in the normal (inflow)
  direction with distance from the reconnection site, which would not
  produce thin SADs as seen in observations.  On the contrary, outflow
  jets in spatially localized three-dimensional reconnection with an
  out-of-plane (guide) magnetic field expand primarily in the
  out-of-plane direction and remain collimated in the normal
  direction, which is consistent with observed SADs being thin.
  Two-dimensional proof-of-principle simulations of reconnection with
  an out-of-plane (guide) magnetic field confirm the creation of
  SAD-like depletion regions and the necessity of density
  stratification.  Three-dimensional simulations confirm that
  localized reconnection remains collimated.
\end{abstract}





\section{INTRODUCTION}
\label{sec-intro}

Supra-arcade downflows (SADs) are dark features at the tops of coronal
arcades that descend sunward during solar flares
\citep{McKenzie99,McKenzie00}.  They are important because they
provide a window into the spatiotemporal evolution of the magnetic
reconnection process that releases energy by changing magnetic
topology during solar flares.  For example, the standard model of
solar flares [{\it e.g.},
\citep{Carmichael64,Sturrock66,Hirayama74,Kopp76}] depicts a single
monolithic reconnection site, but the burstiness and patchiness of
SADs suggests that reconnection is patchy and bursty as well.  This
suggests reconnection occurs at multiple sites in an extended region
rather than along a single line.

There have been many observational studies of SADs, also called
``tadpoles'' because of their sinuous shape.  Their darkness is due to
a lack of emitting particles: they persist in soft X-ray and EUV,
corresponding to temperatures of $10^4-10^7$~K \citep{Innes03}.  Their
sunward speed is 45-500~km/s, below typical coronal Alfv\'en speeds of
1000 km/s \citep{McKenzie00}, though they can be considerably faster
\citep{Liu13}.  They decelerate at $\sim$1500 m/s$^{2}$ and last a few
minutes \citep{Sheeley04}.  They are correlated with bursts of hard
X-rays, suggesting an association with reconnection at the flare site
\citep{Asai04}.  They are most easily seen in the decay phase of long
duration events \citep{McKenzie00}, but they begin during the rise
phase and persist during hard X-ray bursts \citep{Khan07}.

Statistical studies of SADs revealed their cross-sectional areas are a
few to 70 times $10^6$~km$^2$ (a radius of 1-4~Mm), their mean flux is
$\sim 10^{18}$~Mx, they release $10^{27}-10^{28}$~erg of energy, and
their densities are 5-43 times lower than ambient densities
\citep{McKenzie09,Savage11}.  They originate at heights of
$100-200$~Mm and penetrate $20-50$~Mm \citep{Savage10,Savage11}.
Their areas have a log-normal distribution, and their fluxes are
consistent with both log-normal and exponential distributions
\citep{McKenzie11}.  Reconnection inflows and outflows near SADs were
recently observed \citep{Savage12b}.

The explanation of their cause remains under debate.  They were
originally thought of as flux tubes contracting under tension after
reconnection \citep{McKenzie99,McKenzie00}.  \citet{Asai04} support
their relation to reconnection outflows.  From the side, the whole
loop should be visible; these SAD loops (SADLs) were observed
\citep{Savage10,Warren11}.  Recently, it was argued that SADs are not
flux tubes themselves, but wakes behind them \citep{Savage12}.

There have only been a few simulation studies of SADs.  Seminal
studies of flux tubes produced by short-lived patchy three-dimensional
(3D) reconnection \citep{Linton06,Longcope10} in the vein of
\citet{Semenov83b} reproduced some aspects of SADs, including a
teardrop-shaped cross section \citep{Linton06}, deceleration as they
reach underlying arcades \citep{Linton09b}, and trailing density voids
\citep{Guidoni11}.  The evolution of retracting flux tubes were also
studied \citep{Shimizu09}.  It was suggested that SADs are generated
by interfering shocks ahead of locations where energy is deposited by
reconnection \citep{Costa09,Schulz10,Maglione11,Cecere12}.

In an avenue of research that may be relevant to SAD statistics, a
Vlasov-type equation for plasmoids predicts an exponential
distribution of sizes \citep{Fermo10,Uzdensky10,Fermo11,Huang12}.  The
distribution of plasmoid sizes in resistive magnetohydrodynamics (MHD)
was exponential, consistent with observations if corrected for
resolution constraints in detecting small SADs \citep{Guo13}.

There are significant unresolved theoretical issues about present
interpretations of SADs.  For example, if SADs are either flux tubes
or wakes behind them, why does the ambient coronal plasma not rapidly
``fill in'' the region behind them?  If SADs are caused by patchy
reconnection, what causes reconnection to start and stop so rapidly?

We argue that SADs are caused by downflowing outflow jets from
reconnection that is localized in space but not time, as opposed to
isolated flux tubes.  New elements of the present model are (1) the
importance of density stratification of the solar corona and (2) the
role of continuous reconnection as opposed to bursty reconnection in
preventing the corona from filling in the region behind SADs.
We perform proof-of-principle 2D simulations that verify this
mechanism produces SAD-like density depletions and show the necessity
of density stratification.  We show in 3D simulations that localized
reconnection is necessary to keep the outflow jets collimated.  The
model is described in Sec.~\ref{sec-model}.  The simulations are
described in Sec.~\ref{sec-sims}, results are in
Sec.~\ref{sec-results}, and a discussion is in Sec.~\ref{sec-disc}.

\section{A MODEL OF SADs}
\label{sec-model}

\begin{figure}
\epsscale{0.75}
\plotone{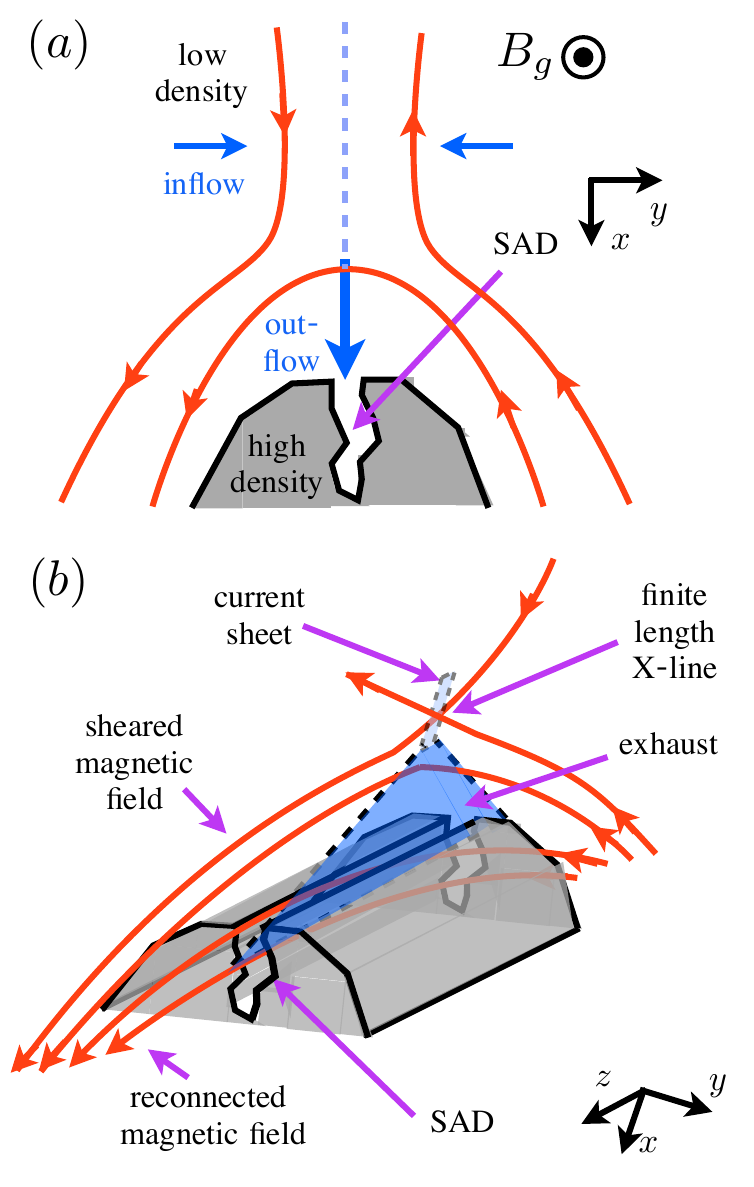}
\caption{\label{fig-model} (a) Schematic diagram of the model of SADs.
  Red lines are magnetic fields.  The vertical dashed blue line is the
  reconnection site high in the corona with a lower density plasma.
  The blue arrows are the reconnection inflow and sunward outflow.
  The SAD is caused by the low density outflow impinging on the high
  density arcade (in gray).  (b) Perspective view showing sheared
  magnetic fields with a finite length X-line and reconnected fields
  threading the SAD formed by the exhaust.}
\end{figure}

Consider a coronal arcade with an overlying vertical current sheet,
sketched in Fig.~\ref{fig-model}(a).  When reconnection begins at the
current sheet, sunward and anti-sunward outflow jets are formed; the
sunward jet is shown by the vertical (blue) arrow.  Being relatively
high in the corona, the plasma at the reconnection site has a
relatively low density because the corona is stratified.  The
relatively higher density plasma in the underlying arcade is denoted
by the shaded region.  Since the density scale height for coronal
parameters is $H \simeq 30-60$ Mm, a jet penetrating 10s of Mm sees a
significant density change.  The density contrast can become larger as
chromospheric evaporation populates the arcade.  The low density
plasma penetrates the denser material below, forming a thin region of
lower density plasma.  We argue this density depletion is a SAD.


This model raises a question about the shape of SADs.  In standard
models of 2D steady-state fast reconnection, the outflow jet expands
outward in the normal direction in Petschek-type exhausts
\citep{Petschek64}.  A similar expansion for bursty reconnection was
described as a snow-plow effect \citep{Semenov98} caused by the
retracting flux tube surrounding increasingly more mass.  Due to the
large length scales between the reconnection site and the looptop, one
might expect the broadened jet to create a depletion region wider than
seen in observations.  However, it was argued that the snow-plow
effect does not occur in 3D localized reconnection if there is an
out-of-plane (guide) magnetic field because the reconnection expands
along the guide field instead of the inflow direction, which
accommodates the mass that causes expansion in 2D \citep{Linton06}.
The guide field during flares has been estimated to be comparable to
the reconnecting magnetic field \citep{Qiu09,Qiu10,Shepherd12}, so
this limit is relevant for solar flares.  

Since the region of the corona in question has a low plasma $\beta$
(the ratio of gas pressure to magnetic pressure), the mass is not
expected to control the magnetic structure of the reconnection
exhaust, so we motivate the result of \citet{Linton06} solely in terms
of conservation of magnetic flux.  Consider a recently reconnected
flux tube with flux $\Delta \Phi$ formed by reconnection at an X-line
of finite extent $l_z$ in the out-of-plane direction $z$ and length
$\Delta l_x$ in the outflow direction $x$.  Just downstream of the
X-line,
\begin{equation}
  \Delta \Phi \sim B_y l_z \Delta l_x, \label{phizeqn}
\end{equation}
where $B_y$ is the reconnected (normal) component of the magnetic
field.  
After the flux tube has convected downstream, the magnetic field in
the exhaust is predominantly in the $z$ direction, so the flux is
\begin{equation}
  \Delta \Phi^{\prime} = B l_{y}^{\prime} \Delta l_x, \label{later}
\end{equation}
where $l_y$ is the width of the flux tube in the $y$ direction, $B =
|{\bf B}|$ is the magnitude of the total magnetic field, and the prime
denotes post-convection downstream.  By conservation of flux, $\Delta
\Phi = \Delta \Phi^{\prime}$, so solving Eqs.~(\ref{phizeqn}) and
(\ref{later}) for $l_y^{\prime}$ gives
\begin{equation}
  l_{y}^{\prime} \sim  l_z \frac{B_y}{B}. \label{lyprimeeqn}
\end{equation}
Regardless of whether reconnection is 2D or 3D, the maximum value of
$B_{y}$ is approximately 0.1 of the reconnecting magnetic field $B_x$
since the normalized rate of fast reconnection is close to 0.1.
Equation \ref{lyprimeeqn} shows that the maximum downstream expansion
of the exhaust $l_{y}^{\prime}$ is proportional to $l_z$.  In quasi-2D
systems, $l_z$ is large and the exhaust gets wider with downstream
distance.  In 3D localized reconnection, $l_y^{\prime}$ has an upper
limit, so the exhaust is collimated.  For the present application,
this implies the density depletion regions formed by finite length
X-lines are thin, consistent with observations of SADs.


A sketch with a perspective 3D view is shown in
Fig.~\ref{fig-model}(b), illustrating the difference between the
present model and models with SADs as isolated flux tubes or wakes
behind them \citep{Savage12}.  In the flux tube model, reconnection
happens for a short time, leading to a single flux tube propagating
sunward.  In the present model, the reconnection is continuous in
time, at least long enough for the material ejected sunward to
propagate to the looptop.  Therefore, reconnected magnetic field
threads the entire extent of the depletion region, not just through a
single flux tube.  This explains why plasma does not rapidly fill in
behind the SAD as would occur if it were an isolated flux tube.
Further comparison to previous work is presented in
Sec.~\ref{sec-disc}.

\section{SIMULATIONS}
\label{sec-sims}

Simulations are performed using the two-fluid code {\sc F3D}
\citep{Shay04}.  Magnetic fields and number densities are normalized
to arbitrary values $B_{0}$ and $n_{0}$.  Lengths are normalized to
the ion inertial scale $d_{i,0} = c / \omega_{pi,0}$, the scale where
the Hall effect is important, where $\omega_{pi,0} = (4 \pi n_{0}
e^{2} / m_{i})^{1/2}$ is the plasma frequency, and $e$ and $m_{i}$ are
the ion charge and mass.  Times are normalized to the inverse ion
cyclotron frequency $\Omega_{ci}^{-1}$, velocities to the Alfv\'en
speed, and pressures to $B_{0}^{2}/4 \pi$.  All simulations have
periodic boundary conditions in each direction.

The 2D simulations have $2048 \times 512$ cells in a system of size
$L_{x} \times L_{y} = 102.4 \times 25.6$.  The continuity, momentum,
ion pressure, and induction equations are evolved in time.  The
electric field is given by the generalized Ohm's law (in cgs units)
\begin{equation} 
  {\bf E} = - \frac{{\bf v}_{i} \times {\bf B}}{c} +
  \frac{{\bf J} \times {\bf B}}{nec} + \frac{m_{e}}{e^{2}} 
  \frac{d ({\bf J}/n)}{dt}, \label{genohm}
\end{equation}
where ${\bf E}, {\bf v}_{i}, {\bf B},$ and ${\bf J}$ are the electric
field, ion bulk flow velocity, magnetic field, and current density,
respectively, $n$ is the number density and $c$ is the speed of light.
Electron inertia $m_{e}$ is included with $m_{e} = m_{i}/25$; this
value has been used often and does not affect the reconnection rate
[{\it e.g.,} \citep{Shay98b}].  Ions are assumed to be adiabatic with
a ratio of specific heats $\gamma = 5/3$ and electrons are assumed to
be cold.

The equilibrium consists of a double Harris sheet for the reconnecting
magnetic field $B_{x}$:
\begin{equation}
  B_{x}(y) = 1 + \tanh\left(\frac{y-y_{0}}{w_{0}}\right) - \tanh\left(\frac{y+y_{0}}{w_{0}}\right) \label{harris}
\end{equation}
with an initial current sheet thickness $w_{0} = 1$ and field reversal
locations of $\pm y_{0} = \pm L_{y} / 4$.  The equilibrium density
profile $n(y)$ is
\begin{equation}
  n(y) = 1 + (n_{p}-1) \left[{\rm sech}^{2}\left(\frac{y - y_{0}}{w_{0}}\right) + {\rm sech}^{2}\left(\frac{y + y_{0}}{w_{0}}\right)\right], \label{density}
\end{equation}
which is 1 far from the current sheets and $n_{p}$ at $y = \pm y_{0}$.
The temperature is initially uniform at $T = 1$.  To balance total
pressure, a guide field $B_{z}$ is employed with $B_{z} = 3$
asymptotically far from the current sheets and a profile that balances
total pressure.  Reconnection is initiated by a magnetic perturbation
$\delta{\bf B} = -(0.012 B_{0} L_{y} / 2 \pi) \hat{{\bf z}} \times
\nabla [\sin(2 \pi x/ L_{x}) \sin^{2}(2 \pi y/L_{y})]$, and a very
small incoherent magnetic perturbation is used to break symmetry.

The 3D simulations have a
system size of $51.2\times 25.6 \times 256$ with $1024 \times 512
\times 256$ grid cells.  The equilibrium is the same as in 2D with
$n_{p} = 1.5$ and $w_{0} = 0.2$.  For this choice of $n_{p}$, a
uniform guide field $B_{z} = 3$ enforces pressure balance.  The ions
are isothermal in the 3D simulations.

Since reconnection with a guide field and the Hall and electron
inertia terms in Eq.~(\ref{genohm}) naturally spreads in the
out-of-plane direction \citep{Shepherd12}, we omit these terms but
include a localized (anomalous) resistivity $\eta_{anom}$ to achieve
fast reconnection of finite extent.  The form of $\eta_{anom}$ is
\begin{equation}
  \eta_{anom} = \eta_{0} e^{-(x^2+y^2)/w^2}\frac{[\tanh(z+w_{z}) - 
    \tanh(z-w_{z})]}{2}, \label{etaanom}
\end{equation}
where $\eta_{0} = 0.01$, $w = 0.5$, and $w_{z}$ is a parameter
controlling the half-length of the reconnection X-line in the
out-of-plane direction.

This equilibrium should be treated solely as a numerical device to set
up a stratified plasma in the outflow direction.  The initial
conditions are stratified in the inflow direction, but when
reconnection begins, it processes this material and populates the
arcade with the high density plasma.  When the lower density material
convects to the reconnection site, the outflow jet is ejected into a
vertically stratified arcade as desired.

\section{RESULTS}
\label{sec-results}

\begin{figure}
\epsscale{1}
\plotone{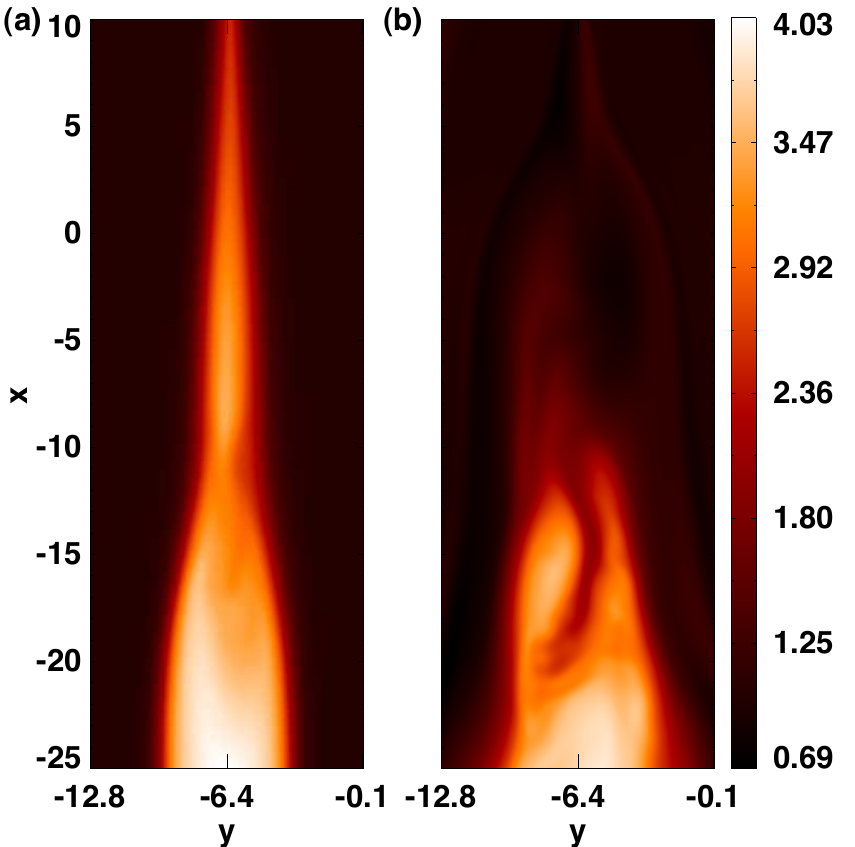}
\caption{\label{fig-2dplots} Plasma density for the $n_{p} = 4$
  simulation at (a) $t = 150$ and (b) $t = 210$.  The presence of a
  density depletion is clear in (b).}
\end{figure}

First, we show that SAD-like depletions occur using $n_{p} = 4$ for
the peak density.  As time evolves, the high density plasma initially
at the current sheet is corralled into the arcade by reconnection
early in the simulation [Fig.~\ref{fig-2dplots}(a) at $t = 150$].
When the lower density plasma reaches the reconnection site, it
populates the outflow jet and penetrates the high density arcade
[Fig.~\ref{fig-2dplots}(b) at $t = 210$].  A sinuous plasma depletion
is clearly seen.  Therefore, reconnection can carve a density
depletion into a higher density arcade.

\begin{figure}
\plotone{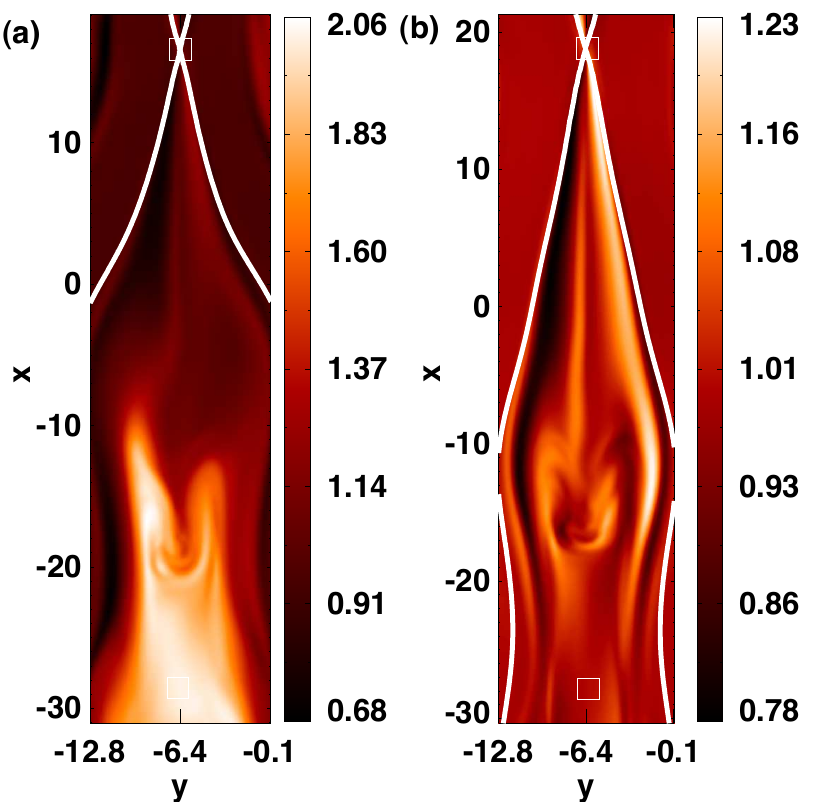}
\caption{\label{fig-density} Plasma density for simulations with
  initial peak density $n_{p}$ of (a) 2 and (b) 1, showing that
  SAD-like depletions require density stratification to occur.}
\end{figure}

To test the importance of density stratification, simulations with
different $n_{p}$ are performed.  Figure~\ref{fig-density}(a) shows
the density in a simulation with $n_{p} = 2$ instead of 4.  The white
line marks the separatrix (the outer-most closed field line), and the
white boxes at the top and bottom denote the X- and O-line,
respectively.
A SAD-like depletion is clearly present.  Panel (b) has an initially
uniform density ($n_{p} = 1$).  While a curved structure does appear,
it is not a depletion as for $n_{p} > 1$.  Therefore, the simulations
suggest that density stratification is a key ingredient of SAD-like
depletions.

Based on the results of Fig.~\ref{fig-2dplots}, one might conclude
that a 2D model with density stratification is sufficient to explain
SADs.  However, we argue there are two reasons the density depletion
cannot propagate the large distances required to explain SADs in 2D.
First, as discussed in Sec.~\ref{sec-model}, fast reconnection in 2D
has an exhaust that expands as in the Petschek model, which is
inconsistent with the collimated structure of SADs.  Second, the
depletion region in Fig.~\ref{fig-2dplots}(b) stretches the
reconnected field so that the field wraps around the front of the jet,
producing a backward tension force that slows the jet.  In 3D
reconnection with a finite length X-line, both of these problems are
fixed.

\begin{figure}
\plotone{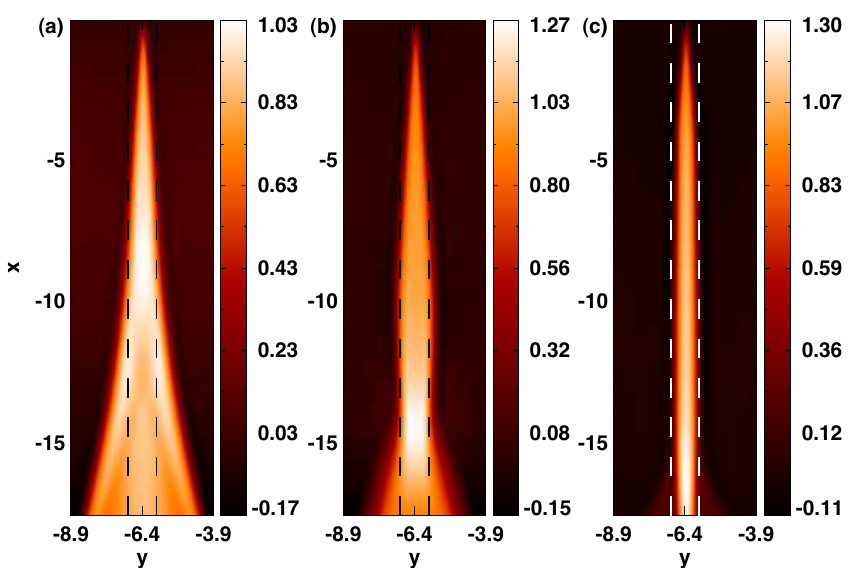}
\caption{\label{fig-locresoutflow} Results for 3D simulations with
  anomalous resistivity localized in the out-of-plane direction.  The
  outflow $v_{x}$ is shown for $w_{z}$ = (a) 60, (b) 30, and (c) 10,
  revealing that the jet becomes increasingly collimated for
  reconnection sites that are more localized in the out-of-plane
  direction.}
\end{figure}

\begin{figure}
\epsscale{0.65}
 \plotone{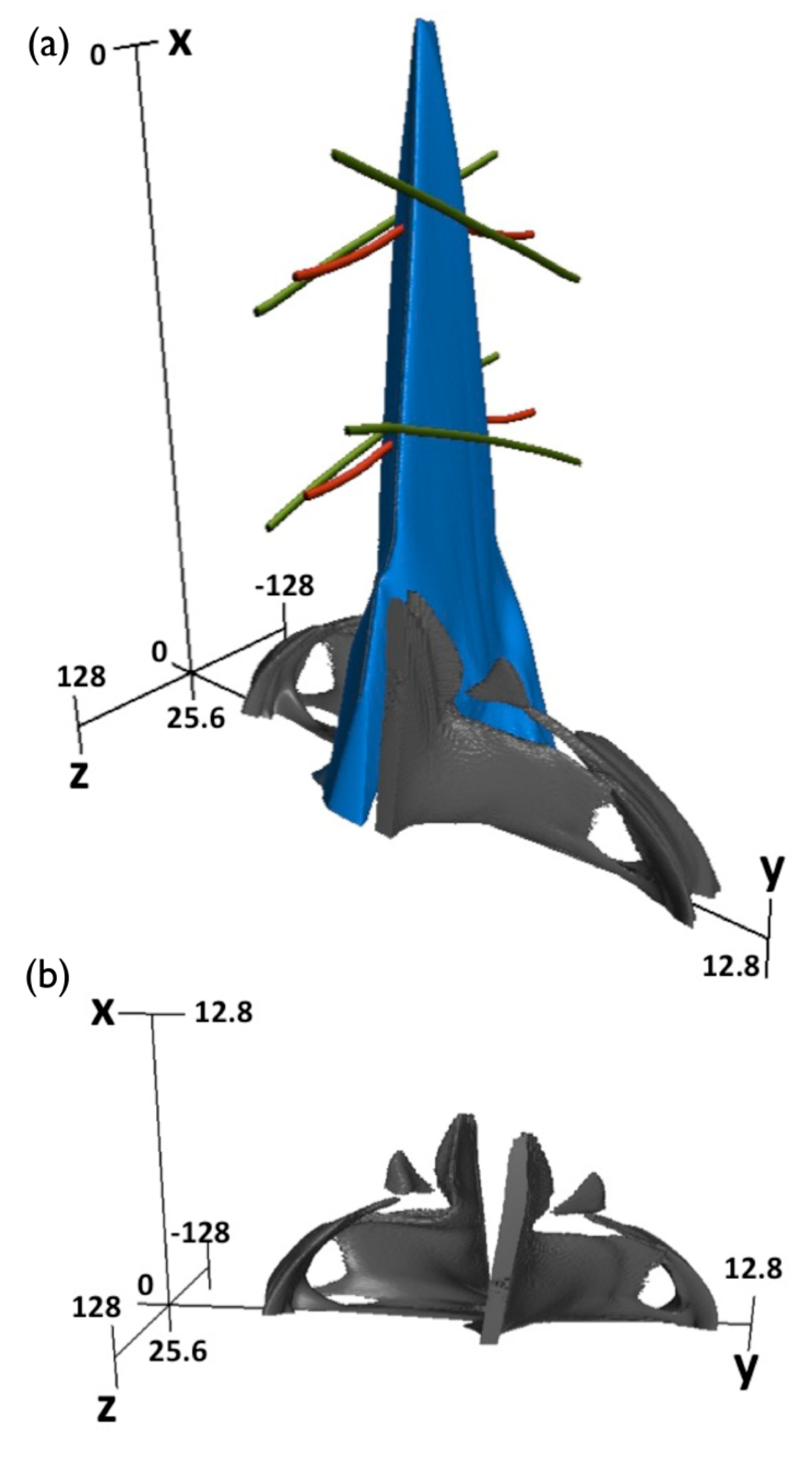}
 \caption{\label{fig-3d} (a) 3D view from the $w_{z} = 10$ simulation
   of isosurfaces of the outflow speed (in blue) showing the exhaust
   spreading in $z$ but not $y$ and a relatively high value of density
   (gray) showing the depletion carved into the arcade.
   Representative unreconnected (red) and reconnected (green) magnetic
   field lines are shown.  (b) The density isosurface from panel (a)
   rotated to reveal the 3D SAD-like density depletion.}
\end{figure}

To see that outflow jets in 2D fast reconnection broaden with distance
from the X-line while localized 3D reconnection leads to collimated
exhausts as seen in the observations, we present 3D simulations with
$n_{p} = 1.5$.  The reconnection outflow $v_{x}$ is shown in
Fig.~\ref{fig-locresoutflow}(a) through (c) for $w_{z} = 60, 30,$ and
$10$, respectively.  The vertical lines are a distance 0.5 from the
neutral line.  The jet broadens continuously for the longest X-line
simulation in panel (a), which for the present computational domain is
essentially a 2D case.  (Note, this is true because of the finite size
of the computational domain; 3D effects would be seen if the system
was larger in the $x$ direction.)  The width of the exhaust remains
limited when reconnection is more localized in the $z$ direction as in
panel (c).  A 3D visualization of the exhaust jets and threading of
the magnetic fields for the $w_{z} = 10$ simulation is given in
Fig.~\ref{fig-3d}(a), which shows that the exhaust is localized in the
$y$ direction and expands in the $z$ direction.  Unlike in the 2D
simulations, the reconnected field lines in red threading the exhaust
are straight (which is obscured in the figure but can be seen when the
exhaust isosurface in blue is removed), which implies there is no
backward tension force slowing the exhaust in 3D.  Panel (b) shows the
same data with the exhaust removed and rotated to more clearly see the
3D density depletion in the underlying arcade.  Thus, 3D spatial
localization of reconnection causes exhaust collimation, a necessary
ingredient to explain SADs.


\section{DISCUSSION}
\label{sec-disc}

We have presented a model with
the critical physics for SAD formation being (1) density
stratification of the corona so that the outflow jet can carve out a
depletion, (2) reconnection being continuous instead of bursty to
prevent the coronal plasma from filling in behind the SAD, and (3)
localization of the reconnection site in the out-of-plane direction in
order for the outflow jet to remain collimated over the large
distances.
Proof-of-principle simulations confirm these three basic
features.



The present model differs from previous ones in a number of important
ways and has a number of appealing aspects because it potentially
answers open questions about SADs.  Previous models suggested SADs are
a result of spatially localized and temporally bursty reconnection.
It is not clear why reconnection stops abruptly.  In the present
model, SADs are localized in space but not in time.  We argue the
observed burstiness of SADs arises from the abrupt onset of
reconnection in disparate spatially localized regions.

A very important question about the interpretation of SADs as flux
tubes or wakes behind them is why the high density corona does not
rapidly fill in behind the descending flux tube.  If it did, SADs
would appear as a descending circle instead of an extended depletion.
In our model, elongated depletions are a natural consequence of their
formation as collimated reconnection outflow jets.
An alternate explanation within the flux tube model has been explored
in terms of peristaltic flow \citep{Scott13}.

\citet{Asai04} previously discussed the correlation of SADs with
reconnection outflows, but no clear distinction was made between
outflow jets and plasmoids ejected from the reconnection site.  There
are outward similarities between the present model and one by
\citet{Costa09,Schulz10,Maglione11,Cecere12}.  These studies used a
pressure pulse to emulate the effect of energy deposition from
reconnection.  It was argued that shocks and waves generated from the
pulse interfere to produce voids.  This occurs even in a uniform
plasma, while here the reconnection jet itself carves the SAD and
requires density stratification.  With the exception of
\citet{Cecere12}, we know of no previous simulation study to include
density stratification.


The present model has similarities to a leading model for low density
plasma bubbles propagating earthward in the Earth's magnetotail
\citep{Pontius90}, though there are some differences.  In both models,
reconnection jets impinge on a higher density region underlying the
reconnection site.  In the magnetotail, an extended reconnection
X-line in the out-of-plane direction forms and breaks up due to
interchange into numerous flow channels \citep{Wiltberger00}.  Here,
we argue the X-line is spatially localized, leading to a single flow
channel.

Of course, SAD-like depletions in simulations should not be overstated
as real SADs.  The many similarities to observed SADs are qualitative
rather than quantitative.  Future work will require quantitatively
investigating whether the following properties in the simulations
agree with observations: sub-Alfv\'enic sunward speed, deceleration,
duration, penetration depth, sizes, and energy content.  

The authors gratefully acknowledge support from NSF grants AGS-0953463
(PAC), AGS-1202330 (JFD), ATM-0645271 (MAS) and NASA grants NNX10AN08A
(PAC), \\ NNX10AC01G (JTG), NNX08AO83G (TDP), and NNX08AO84G (JTG) and
acknowledge beneficial conversations with J.~T.~Karpen, W.~Liu,
J.~C.~Raymond, and M.~I.~Sitnov.  This research used resources of the
National Energy Research Scientific Computing Center.



\clearpage

\clearpage

\clearpage

\clearpage

\clearpage

\end{document}